\documentclass[journal]{IEEEtran}
\IEEEoverridecommandlockouts

\usepackage{silence}
\usepackage{cite}

\usepackage{amsmath,amssymb,amsthm,fixmath}

\usepackage{color,colortbl}
\usepackage{xcolor}

\usepackage[inline]{enumitem}

\usepackage{siunitx}
\DeclareSIUnit{\dBm}{dBm}

\usepackage{graphicx}
\graphicspath{{simulations/}}
\usepackage[labelformat=simple]{subcaption}

\usepackage{multirow,booktabs}

\usepackage{optidef}

\usepackage[hidelinks]{hyperref}

\usepackage[shortcuts,acronym,automake]{glossaries}
\makeglossaries
\newacronym{awgn}{AWGN}{additive white Gaussian noise}
\newacronym{blcp}{BLCP}{block confusion probability}
\newacronym{blep}{BLEP}{block erasure probability}
\newacronym{fap}{FAP}{false alarm probability}
\newacronym{bler}{BLER}{block error rate}
\newacronym{blec}{BLEC}{block erasure channel}
\newacronym{cdf}{CDF}{cumulative distribution function}
\newacronym{fbl}{FBL}{finite blocklength}
\newacronym{ibl}{IBL}{infinite blocklength}
\newacronym{kld}{KLD}{Kullback--Leibler divergence}
\newacronym{lp}{LP}{linear program}
\newacronym{swhd}{SWHD}{symbol-wise hard-decision decoder}
\newacronym{lrt}{LRT}{likelihood-ratio test}
\newacronym{pdf}{PDF}{probability density function}
\newacronym{psk}{PSK}{phase-shift keying}
\newacronym{snr}{SNR}{signal-to-noise ratio}
\newacronym{urllc}{URLLC}{ultra-reliable low-latency communication}
\newacronym{csi}{CSI}{channel state information}
\newacronym{drl}{DRL}{deep reinforcement learning}
\newacronym{gpu}{GPU}{graphics processing unit}
\newacronym{ip}{IP}{Internet Protocol}
\newacronym{kl}{KL}{Kullback-Leibler}
\newacronym{ldpc}{LDPC}{low-density parity-check}
\newacronym{mac}{MAC}{message authentication code}
\newacronym{map}{MAP}{maximum a posteriori}
\newacronym{mdp}{MDP}{Markov decision process}
\newacronym{ml}{ML}{maximum likelihood}
\newacronym{pid}{PID}{proportional-integral-derivative}
\newacronym{ppo}{PPO}{proximal policy optimization}
\newacronym{qpsk}{QPSK}{quadrature phase-shift keying}

\theoremstyle{definition}
\newtheorem{remark}{Remark}

\usepackage[norelsize,linesnumbered,ruled]{algorithm2e}

\newcommand{\diff}{\text{d}}

\usepackage[normalem]{ulem}

\usepackage[textsize=tiny,colorinlistoftodos]{todonotes}
\makeatletter
\define@key{todonotes}{bh}[]{%
	\setkeys{todonotes}{author=\textbf{Bin}, color=lime!30}}%
\define@key{todonotes}{yz}[]{%
	\setkeys{todonotes}{author=\textbf{Yao}, color=blue!30}}%
\makeatother

\newif\ifreviewmode
\reviewmodefalse

\ifreviewmode
\else
  \renewcommand{\todo}[1]{}
  
\fi

\begin{document}

\title{Litter-Masked Block z-Channel: Designing Cover Distributions Against Public-Design Observers}

\author{
	\IEEEauthorblockN{
		Bin~Han,~\IEEEmembership{Senior Member,~IEEE,}
		Wenwen~Chen,
        Yao~Zhu,~\IEEEmembership{Member,~IEEE,}\\
		Rafael~F.~Schaefer,~\IEEEmembership{Senior Member,~IEEE,}
		Giuseppe~Caire,~\IEEEmembership{Fellow,~IEEE,}
        and~Hans~D.~Schotten,~\IEEEmembership{Member,~IEEE}
	}%
	\thanks{B. Han, W. Chen, and H. D. Schotten are with RPTU University Kaiserslautern-Landau. Y. Zhu is with Wuhan University.  R. F. Schaefer is with TU Dresden. G. Caire is with TU Berlin. H. D. Schotten is with the German Research Center for Artificial Intelligence (DFKI). B. Han (bin.han@rptu.de) and Y. Zhu (yao.zhu@whu.edu.cn) are the corresponding authors.}
}

\maketitle

\pagestyle{empty}
\thispagestyle{empty}

\begin{abstract}
Networking defenses against traffic analysis fill idle slots with cover traffic, yet the physical layer carrying it still hands upper layers the erasure-only model of a silent-idle link, leaving the cover content unexploited. We shape that content: the idle-slot waveform is drawn from the codebook complement under a design distribution, the \emph{litter distribution}, chosen to defeat a public-design passive observer while the upper layer keeps its block z-channel model. The \ac{map} receiver reduces to a single threshold on the log-likelihood ratio between the best codeword and the prior-weighted litter aggregate; its errors are silent corruptions the upper layer cannot detect and erasures that retransmission absorbs. Receiver design is thus one-dimensional, minimizing erasure under a cap on silent corruption, and the joint design a Stackelberg game against a \ac{kl} indistinguishability metric, solved by an alternating convex relaxation and corroborated by an independent reinforcement-learning policy. On a rate-$3/4$ $(16,12)$ \ac{ldpc} code against an observer who knows her own channel but whose channel the transmitter knows only statistically, shaped litter cuts the observer's Stein detection exponent by $41$ to $48\%$ relative to uniform cover, flat over a $10$~dB span and in the activity rate. Below the point where the legitimate link turns reliable, no cover distribution admits an operating point; at that point shaping costs about one percent of throughput, and nothing measurable $2$~dB above it.
\end{abstract}


\glsresetall

\section{Introduction}\label{sec:intro}

A transmitter that goes silent between transmissions tells a passive observer when its operator is active. Networking defenses against the resulting traffic-analysis and activity-inference attacks fill the idle transmit slots with \emph{cover traffic} that mimics, on the medium, the bursts the transmitter sends when it does have something to say~\cite{WCM2009morphing,DCRS2012peekaboo,JIPDW2016wtfpad,SRM+2013meter,Apthorpe2017smarthome,ASNF2019stp,TK2021countermeasure}. The physical layer that carries those bursts and that cover, however, presents the upper layers with a small set of error abstractions, and the dominant one is the \emph{packet erasure}~\cite{BLM+1998digital,HMK+2006networkcoding}: delivered packets are authentic, in the sense that codeword confusions are negligible enough for the upper layer to ignore them by design, and missing packets are erasures that automatic repeat request~\cite{LCM1984arq} or rateless fountain codes~\cite{Luby2002lt,Shokrollahi2006raptor} recover. Under a silent-idle convention this strengthens to a \emph{block z-channel} model by adding a third property: idle slots cannot inject phantom packets into the receive queue, because the silent-idle null falls outside any codeword's decision region at any reasonable \ac{snr}. Our recent work validated the packet-erasure abstraction itself at finite blocklength~\cite{HZS+2026confusions}; a follow-on preprint~\cite{Han2026confusions} extended the validation to the block z-channel model by showing that the false-alarm term vanishes under the silent-idle null. Cover traffic removes the silent-idle null on which that extension rests.

This paper replaces the silent-idle null with a \emph{litter sequence} drawn from the codebook complement under a design distribution we call the \emph{litter distribution}, and treats the resulting cover-on-complement system as a \emph{mechanism design problem} with two coupled levels: the receiver sets its decoder threshold to minimize the erasure rate, subject to a cap on the rate at which the link delivers a phantom valid-looking packet to the upper layer; the transmitter shapes the litter distribution to minimize a public-design passive observer's power to tell idle from active slots, subject to the erasure the receiver is then left with staying within the retransmission budget. The intellectual ancestor of the construction is Rivest's chaffing-and-winnowing~\cite{Rivest1998chaffing,BB2000security}: authenticated wheat packets interleaved with fake chaff packets that the receiver discards by a \ac{mac} tag. The litter-masked block z-channel is the physical-layer analogue, separating codewords from litter by codebook membership rather than by an authentication tag, and inheriting the cover-distribution-shaping paradigm that the networking-privacy community has developed at the \ac{ip} layer, from traffic morphing as convex programming~\cite{WCM2009morphing} to stochastic traffic padding~\cite{Apthorpe2017smarthome,ASNF2019stp}, but applying it per block at the physical layer, which this line of work has not yet reached.

The litter-masked block z-channel admits two coupled design axes. The receiver-side axis asks how to draw the codeword-versus-litter line at fixed litter distribution, and reduces to a one-dimensional choice of a single threshold on the log-likelihood ratio between the best-codeword likelihood and the prior-mass-weighted litter aggregate. The transmitter-side axis asks how to shape the litter distribution so the idle-slot waveform is indistinguishable, in a \ac{kl} sense, from active transmissions to a public-design passive observer. The two axes couple: the receiver's optimal threshold depends on the litter distribution through the log-likelihood-ratio statistic, and the indistinguishability constraint must be satisfied at whatever threshold the receiver picks. We frame the joint problem as a Stackelberg game between the transmitter's litter shaping (leader) and the receiver's threshold response (follower), characterize when the inner problem is convex, and solve the non-convex joint case by an alternating relaxation backed by a constrained reinforcement-learning policy for the regime where a legitimate-link distinguishability floor binds.

Four main lines of work have addressed problems closely related to the one posed here: (i)~cover-traffic shaping in the networking-privacy community~\cite{WCM2009morphing,Apthorpe2017smarthome,ASNF2019stp,AHRNF2019stp,XSM2022shaping,SVGSLM2024netshaper,HCOH2024detorrent}, which treats the selection of cover traffic as a design-and-optimization problem at the \ac{ip}, transport, and \ac{mac} layers; (ii)~joint detection and decoding~\cite{WM2014codeword,LOD2021joint,TCW2009async}, which analyzes the binary active-versus-idle decision together with the decoding of the active block; (iii)~covert communication~\cite{BGT2013limits,WWZ2016fundamental,Bloch2016covert,TB2019first}, which characterizes achievable rates under an indistinguishability constraint, with keyless variants~\cite{SBG+2017jammer,GBG+2016warden} that replace the shared key by auxiliary randomness; and (iv)~physical-layer cover traffic~\cite{TK2021countermeasure,DRM2019pnask,BDRBM2021stealte} and constellation shaping for covertness~\cite{MZS+2022shaping}, which embed the cover inside the active waveform. All four stop short of the problem addressed here: a designable distribution on the codebook complement, decoded by a \ac{map} receiver whose threshold is tuned to the three link-layer error events (codeword confusion, erasure, and false alarm), constrained to be indistinguishable to an observer who knows the design, and optimized jointly with the receiver threshold as a Stackelberg game. Line~(i) operates at packet or symbol level, never per block on the codebook complement; line~(ii) assumes a silent or noise-only idle slot throughout; line~(iii) makes the codebook itself the covert carrier and pays for covertness with a shared key or with randomness external to the transmitter-receiver pair; line~(iv) modifies the active waveform rather than the idle one. Two further results enter as tools rather than as competing approaches: Cachin's information-theoretic steganography~\cite{Cachin2004information} supplies the \ac{kl} indistinguishability metric we adopt against the observer, and channel resolvability~\cite{HV1993approximation,Hayashi2006general} together with probabilistic amplitude shaping~\cite{BSS2015bandwidth} construct output-mimicking distributions for an unconstrained input alphabet, whereas here the input alphabet is constrained to the codebook complement.

Our main contributions are fourfold. First, we formulate the litter-masked block z-channel construction and the associated mechanism design problem, with an operational event taxonomy that lists the six elementary transmit-to-receive transitions and absorbs within-litter relabeling into correct idleness, yielding three design metrics: codeword confusion, erasure, and false alarm. Second, we derive the Bayes-optimal \ac{map} decoder for the codeword-plus-litter output space and identify the Bayes-optimal threshold $\tau_{\mathrm{Bayes}}$ in closed form; we then replace the Bayes-optimal threshold by the operational choice that minimizes the erasure rate under a cap on silent corruption (confusion plus false alarm), matching the systems-engineering view in which only detected losses are upper-layer-recoverable. Third, we fix the threat model and indistinguishability metric against a public-design passive observer using a \ac{kl} divergence at the channel output, with an explicit account of what each of the three parties knows about the two links. Fourth, we formulate the joint litter-shaping problem as a Stackelberg game, characterize the convex case, give an alternating algorithm for the non-convex joint case with an independent reinforcement-learning solver as a cross-check, and evaluate the design on a rate-$3/4$ $(16,12)$ \ac{ldpc} instance over $165$ independent runs, where shaped cover removes $41$ to $48\%$ of the observer's detection exponent wherever the legitimate link has an operating point at all.

The remainder of the paper is organized as follows. Sec.~\ref{sec:related_work} places the construction among its neighbors. Sec.~\ref{sec:model} defines the system model, the event taxonomy and the operational metrics. Sec.~\ref{sec:receiver} derives the \ac{map} decoder and its threshold. Sec.~\ref{sec:threat} states the threat model, the knowledge assumed of each party, the indistinguishability metric and the adaptation architecture. Sec.~\ref{sec:shaping} formulates litter shaping as a Stackelberg game and gives the two solvers. Sec.~\ref{sec:numerics} reports the numerical study, and Sec.~\ref{sec:conclusion} concludes.


\section{Related Work}\label{sec:related_work}

The closest \emph{systemic} neighbor is cover-distribution shaping in the networking-privacy community, where cover selection has matured from Rivest's unshaped chaffing-and-winnowing~\cite{Rivest1998chaffing,BB2000security} into a design problem: traffic morphing as a convex program~\cite{WCM2009morphing}, stochastic padding~\cite{ASNF2019stp,AHRNF2019stp}, differentially private shaping~\cite{XSM2022shaping,SVGSLM2024netshaper}, and adversarially trained padding~\cite{HCOH2024detorrent}. All of it acts at packet level, never per-block on a codebook complement, and never couples the cover to a decoder threshold. Physical-layer cover is instead embedded in the \emph{active} waveform: constellation shaping for covertness~\cite{MZS+2022shaping}, wireless steganography~\cite{DSGS2012dirty,DRM2019pnask}, and covert modulation~\cite{TK2021countermeasure}.

The closest \emph{mathematical} neighbor, joint detection and decoding~\cite{WM2014codeword,TCW2009async,LOD2021joint,OCGB2024jdd}, characterizes the decode/false-alarm/missed-detection trade-off to short blocklength, but always with a silent or noise-only idle slot, never a designable waveform on the codebook complement. The covert-communication line~\cite{BGT2013limits,WWZ2016fundamental,Bloch2016covert,TB2019first} makes the codebook itself the covert carrier under shared keys or auxiliary randomness~\cite{SBG+2017jammer,GBG+2016warden}; here the codebook carries the message at full rate and the covertness budget is paid by the idle-slot distribution. Cachin's steganography~\cite{Cachin2004information} supplies our \ac{kl} metric; channel resolvability~\cite{HV1993approximation,Hayashi2006general} and probabilistic amplitude shaping~\cite{BSS2015bandwidth} supply output-mimicry for an unconstrained input alphabet, the very constraint the codebook complement imposes.

\section{System Model}\label{sec:model}

Each slot carries one length-$n$ block of symbols from a finite alphabet $\mathcal{M}$ of size $M\triangleq|\mathcal{M}|$, so the sequence space $\mathcal{M}^n$ holds $M^n$ candidate blocks. A subset $\mathcal{X}\subset\mathcal{M}^n$ with $|\mathcal{X}|=M^k$ is the codebook of an $(n,k)$ block code of rate $R=k/n$, carrying $k\log_2 M$ information bits per active slot. The complement
\begin{equation}
\mathcal{L}\triangleq\mathcal{M}^n\setminus\mathcal{X},\qquad |\mathcal{L}|=M^n-M^k,
\label{eq:litter_set}
\end{equation}
is the \emph{litter set}: the pool of non-codeword sequences from which cover traffic is drawn. Both $\mathcal{X}$ and $\mathcal{L}$ are public knowledge.

\subsection{Transmitter}\label{subsec:tx}

The transmitter puts a length-$n$ block $X\in\mathcal{M}^n$ on the channel every slot, active or silent: $X\in\mathcal{X}$ sends a codeword, $X\in\mathcal{L}$ fills the slot with litter. Activity is Bernoulli with $\mathrm{Pr}\{X\in\mathcal{X}\}=p\in(0,1)$ i.i.d.\ across slots, $p$ an exogenous traffic intensity. Conditioned on $X\in\mathcal{X}$, $P_X$ is uniform ($P_X(x)=M^{-k}$); conditioned on $X\in\mathcal{L}$, the block follows the litter distribution $P_L$, the central design object, whose uniform choice $P_L(\ell)=(M^n-M^k)^{-1}$ is the no-shaping baseline that Sec.~\ref{sec:shaping} optimizes jointly with the receiver.

The two conditionals combine into the unconditional distribution of $X$ on $\mathcal{M}^n$:
\begin{equation}
\mathrm{Pr}\{X=x\}=p\,P_X(x)\mathbf{1}\{x\in\mathcal{X}\}+(1-p)\,P_L(x)\mathbf{1}\{x\in\mathcal{L}\}.
\label{eq:tx_mixture}
\end{equation}

\subsection{Channel and Observation}\label{subsec:channel}

Each symbol $X_t$ maps to a complex constellation point through a unit-average-energy labeling $\mu\colon\mathcal{M}\to\mathcal{S}\subset\mathbb{C}$ The block is received over a channel with complex gain $h$, constant over the block and known at the receiver, in \ac{awgn} of fixed variance $N_0$, the receiver noise floor: $Y'=h\,\mu(X)+Z$ with $Z\sim\mathcal{CN}(0,N_0 I_n)$. Coherent scaling by $h$ gives the observation used throughout,
\begin{equation}
Y=\mu(X)+W,\qquad W\sim\mathcal{CN}(0,\gamma^{-1} I_n),
\label{eq:awgn}
\end{equation}
with per-symbol \ac{snr} $\gamma\triangleq|h|^2/N_0$. For simplicity, the numerical study uses Gray-labeled \ac{qpsk}; the analysis of Secs.~\ref{sec:receiver}--\ref{sec:shaping} holds for any labeling, except that the coset-class construction of Sec.~\ref{subsec:shaping_practical} uses the Gray-\ac{qpsk} geometry. No transmitter--receiver shared secret is assumed. The legitimate and observer links both have this form and differ in their gains, introduced in Sec.~\ref{sec:threat}.

\subsection{Operational Event Taxonomy}\label{subsec:events}

The receiver maps each observed block $Y$ to an output $\hat X\in\mathcal{M}^n$ that takes values in the same sequence space as the transmitted block. Crossing the binary transmit class $\{X\in\mathcal{X},\,X\in\mathcal{L}\}$ with the binary output class $\{\hat X\in\mathcal{X},\,\hat X\in\mathcal{L}\}$ and then resolving each class to specific elements yields six elementary transmit-to-receive transitions:
\textbf{E1} correct decoding ($X=x_i$, $\hat X=x_i$); \textbf{E2} correct idleness ($X=\ell_i$, $\hat X=\ell_i$); \textbf{E3} codeword confusion ($X=x_i$, $\hat X=x_j$, $j\neq i$); \textbf{E4} litter confusion ($X=\ell_i$, $\hat X=\ell_j$, $j\neq i$); \textbf{E5} erasure ($X\in\mathcal{X}$, $\hat X\in\mathcal{L}$); and \textbf{E6} false alarm ($X\in\mathcal{L}$, $\hat X\in\mathcal{X}$).
At the link interface the receiver's output is consumed by the upper layer either as a specific codeword identifier or as ``no codeword sent''; within-litter relabeling never reaches any higher-layer mechanism, by construction of the litter set as the codebook complement. Event \textbf{E4} is therefore operationally indistinguishable from \textbf{E2} and is merged into the correct-idleness outcome at the metric level. The resulting five operational outcomes are correct decoding (\textbf{E1}), correct idleness (\textbf{E2}$\cup$\textbf{E4}), codeword confusion (\textbf{E3}), erasure (\textbf{E5}), and false alarm (\textbf{E6}).

\subsection{Operational Error Metrics}\label{subsec:metrics}

Three conditional probabilities, all defined at the class level, summarize the link-layer behavior:
\begin{align}
P_{\mathrm{con}} &\triangleq \mathrm{Pr}\{\hat X\in\mathcal{X}\setminus\{X\}\mid X\in\mathcal{X}\},\label{eq:Pcon}\\
P_{\mathrm{ers}} &\triangleq \mathrm{Pr}\{\hat X\in\mathcal{L}\mid X\in\mathcal{X}\},\label{eq:Pers}\\
P_{\mathrm{fa}}  &\triangleq \mathrm{Pr}\{\hat X\in\mathcal{X}\mid X\in\mathcal{L}\}.\label{eq:Pfa}
\end{align}
$P_{\mathrm{con}}$ captures the rate of event \textbf{E3} and is the rate at which the link injects a phantom yet syntactically valid packet into the upper-layer stack; $P_{\mathrm{ers}}$ captures the rate of \textbf{E5} and is the rate at which a transmitted packet fails to be delivered; $P_{\mathrm{fa}}$ captures the rate of \textbf{E6} and is the rate at which the link manufactures a packet out of cover traffic. The metric definitions are insensitive to within-litter precision, in line with the merge of \textbf{E4} into correct idleness.

The three metrics play different roles at the link interface. An erasure is a \emph{detected loss}: the upper layer notices the missing block and repairs it by retransmission or outer erasure coding, up to a tolerance $\bar P_{\mathrm{ers}}$ fixed by its retransmission budget. Codeword confusion and false alarm are \emph{silent corruptions}: the link layer delivers a valid-looking packet with no indication that anything went wrong, so link-layer retransmission is never triggered; recovery is possible only at the application layer, by an outer code that treats a confusion as a substitution and a false alarm as an insertion, as Remark~\ref{rem:blockchannel} makes precise, at a cost in outer rate that this paper does not optimize. What the upper layer sees of the two silent events is their combined rate
\begin{equation}
P_{\mathrm{silent}} \triangleq p\,P_{\mathrm{con}} + (1-p)\,P_{\mathrm{fa}},
\label{eq:Psilent}
\end{equation}
the expected per-block rate at which a phantom valid-looking packet enters the upper-layer stack; the block z-channel model of Sec.~\ref{sec:intro} holds as long as $P_{\mathrm{silent}}$ stays below a target $\bar P_{\mathrm{silent}}$. The design problem of this paper follows, and Secs.~\ref{subsec:np_reading} and~\ref{sec:shaping} formalize its two levels: the receiver chooses its threshold to minimize $P_{\mathrm{ers}}$ subject to $P_{\mathrm{silent}}\le\bar P_{\mathrm{silent}}$, and the transmitter shapes the litter distribution subject to the erasure rate the receiver is then left with staying within the retransmission budget $\bar P_{\mathrm{ers}}$. The silent-corruption cap is the hard constraint because a silent corruption is the event the upper layer cannot see; the erasure budget is the price paid for it. The aggregate block error rate
\begin{equation}
P_{\mathrm{err}}=p\,\bigl(P_{\mathrm{con}}+P_{\mathrm{ers}}\bigr)+(1-p)\,P_{\mathrm{fa}},
\label{eq:Perr_total}
\end{equation}
which a Bayes-optimal receiver would minimize, treats a detected loss and a silent corruption alike and is not the objective.

\begin{remark}[Block-level channel seen by the upper layer]\label{rem:blockchannel}
The three metrics describe, from the upper layer's viewpoint, a block channel with substitutions (\textbf{E3}), losses (\textbf{E5}), and insertions (\textbf{E6}). When packets carry sequence numbers, a loss is detected as a gap and becomes an erasure of known position, and an inserted packet either carries an invalid index and is discarded or collides with a legitimate index and becomes a substitution; an outer $(N,K)$ maximum-distance-separable code, such as a Reed--Solomon code with minimum distance $N-K+1$~\cite{Singleton1964mds}, then corrects any $r$ substitutions and $e$ erasures with $2r+e\le N-K$~\cite{Forney1966gmd}. Without such indexing the upper layer faces a channel with synchronization errors, for which capacity and code constructions are far less complete~\cite{Mitzenmacher2009deletion,MBT2010syncsurvey,CR2021synccapacity}. Maximizing the outer code rate over all three metrics jointly with the litter shaping of Sec.~\ref{sec:shaping}, in place of the cap on $P_{\mathrm{silent}}$ adopted here, is a natural extension that we leave open.
\end{remark}

\section{Optimal MAP Decoder}\label{sec:receiver}

The receiver knows the codebook $\mathcal{X}$, the litter set $\mathcal{L}$, the activity prior $p$, and the litter distribution $P_L$, and observes $Y$ from \eqref{eq:awgn}. Its \ac{map} decoder is Bayes-optimal under the zero-one loss on the merged outcomes of Sec.~\ref{subsec:events}, and it collapses to a single threshold on the log-likelihood ratio between the best codeword and the prior-weighted litter aggregate. We keep that threshold as an explicit design parameter rather than fixing it at its Bayes-optimal value, so that the false-alarm and erasure rates can be traded against each other under the caps of Sec.~\ref{subsec:metrics}.

\subsection{Likelihood Statistics}\label{subsec:llrs}

Let $f_s(y)\triangleq p_{Y\mid X}(y\mid s)$ denote the conditional density of receiving $y$ given that $s\in\mathcal{M}^n$ was transmitted; viewed as a function of $s$ at the observed $y$, it is the likelihood of $s$. Under \eqref{eq:awgn},
\begin{equation}
f_s(y)=(\gamma/\pi)^{n}\exp\!\bigl(-\gamma\|y-\mu(s)\|^2\bigr).
\label{eq:awgn_density}
\end{equation}
The codeword-wise \ac{ml} inner decoder
\begin{equation}
\hat x(y)\triangleq\arg\max_{x\in\mathcal{X}}\,f_x(y)
\label{eq:cwml}
\end{equation}
selects the codeword closest to $y$ in Euclidean distance; ties have measure zero under \ac{awgn} and are broken arbitrarily. With this inner step in place we define the best-codeword likelihood
\begin{equation}
L_{\mathcal{X}}(y)\triangleq\max_{x\in\mathcal{X}}\,f_x(y)=f_{\hat x(y)}(y),
\label{eq:LX}
\end{equation}
the aggregated litter likelihood
\begin{equation}
L_{\mathcal{L}}(y)\triangleq\sum_{\ell\in\mathcal{L}}P_L(\ell)\,f_\ell(y),
\label{eq:LL}
\end{equation}
and the log-likelihood-ratio statistic
\begin{equation}
\Lambda(y)\triangleq\log\bigl(L_{\mathcal{X}}(y)/L_{\mathcal{L}}(y)\bigr).
\label{eq:Lambda}
\end{equation}
The asymmetry between \eqref{eq:LX} and \eqref{eq:LL}, a maximum over $\mathcal{X}$ against an expectation over $\mathcal{L}$, mirrors the asymmetry of the outcome merge of Sec.~\ref{subsec:events}: codewords are addressed individually, whereas litter outcomes are merged at the class level.

\subsection{Bayes-Optimal Decision Rule}\label{subsec:bayes_rule}

Let $Z(y)\triangleq p\,M^{-k}\sum_{x\in\mathcal{X}}f_x(y)+(1-p)\,L_{\mathcal{L}}(y)$ be the posterior normalizer under \eqref{eq:tx_mixture}, and let the cost of outputting $\hat x$ when $X=x$ be the zero-one loss on the merged outcomes of Sec.~\ref{subsec:events}: $c(\hat x,x)=0$ if $\hat x=x$ or if $\{\hat x,x\}\subset\mathcal{L}$, and $c(\hat x,x)=1$ otherwise. Under this loss, the Bayes risk of outputting a specific codeword $x_j\in\mathcal{X}$ at observation $Y=y$ is
\begin{equation*}
R(x_j\mid y) = 1 - \mathrm{Pr}\{X=x_j\mid y\},
\end{equation*}
since every transmitted $s\neq x_j$ contributes a unit loss (codeword confusion if $s\in\mathcal{X}\setminus\{x_j\}$, false alarm if $s\in\mathcal{L}$). Under the uniform codeword prior, the index that minimizes $R(x_j\mid y)$ is the codeword-\ac{ml} pick \eqref{eq:cwml}, so the minimum risk of a codeword decision is
\begin{equation*}
R\bigl(\hat x(y)\mid y\bigr) = 1 - \frac{p\,M^{-k}\,L_{\mathcal{X}}(y)}{Z(y)}.
\end{equation*}
The Bayes risk of outputting any specific litter $\ell_m\in\mathcal{L}$ is
\begin{equation*}
R(\ell_m\mid y) = \mathrm{Pr}\{X\in\mathcal{X}\mid y\} = \frac{p\,M^{-k}\sum_{x\in\mathcal{X}}f_x(y)}{Z(y)},
\end{equation*}
independent of which litter is output, because the loss is zero for every litter-to-litter output, in line with the merge of \textbf{E4} into correct idleness. The receiver decides for a codeword whenever $R(\hat x(y)\mid y)<R(\ell_m\mid y)$; substituting the normalizer $Z(y)$ and cancelling the common active-mass term reduces this comparison to $(1-p)\,L_{\mathcal{L}}(y)<p\,M^{-k}\,L_{\mathcal{X}}(y)$, i.e., to the threshold test $\Lambda(y)>\log\bigl[(1-p)M^k/p\bigr]$ on \eqref{eq:Lambda}. The Bayes-optimal rule therefore reduces to
\begin{equation}
\hat X=
\begin{cases}
\hat x(y) & \text{if }\Lambda(y)>\tau,\\
\text{any }\ell\in\mathcal{L} & \text{if }\Lambda(y)\leq\tau,
\end{cases}
\label{eq:decision}
\end{equation}
with the Bayes-optimal threshold
\begin{equation}
\tau_{\mathrm{Bayes}}\triangleq\log\frac{(1-p)\,M^k}{p}.
\label{eq:tau_bayes}
\end{equation}
Since $M^k\gg 1$ in any practical setting, $\tau_{\mathrm{Bayes}}$ is large and positive across the operationally interesting range of $p$, biasing the receiver toward the idle decision in proportion to how dominant the litter mass is in the prior.

When the receiver decides for idleness, any specific litter output is admissible. A natural implementation choice is the litter-\ac{map} pick $\hat\ell(y)=\arg\max_{\ell\in\mathcal{L}}P_L(\ell)\,f_\ell(y)$, but its within-litter precision is operationally invisible at the link interface and contributes equally to event \textbf{E2} or \textbf{E4} of Sec.~\ref{subsec:events}, both of which are correct-idleness under the merge.

\begin{remark}[Maximum-likelihood alternatives]\label{rem:ml}
Two prior-blind rules suggest themselves, and neither is adequate. The codeword-\ac{ml} decoder \eqref{eq:cwml} alone, maximizing over $\mathcal{X}$, never declares idleness: every litter slot becomes a false alarm, so it is the inner step of \eqref{eq:decision} rather than a substitute for it. The \ac{ml} decoder over the full sequence space, $\hat s_{\mathrm{ML}}(y)\triangleq\arg\max_{s\in\mathcal{M}^n}f_s(y)$, can declare idleness by returning a litter, but it weighs the best codeword against the single best litter and ignores $p$ and $P_L$. It is Bayes-optimal only under a uniform prior on $\mathcal{M}^n$, whereas under \eqref{eq:tx_mixture} the mass $p$ is spread over $M^k$ codewords and the mass $1-p$ over $M^n-M^k$ litter sequences, prior weights that differ by orders of magnitude. In the notation of \eqref{eq:Lambda} it amounts to replacing the aggregate $L_{\mathcal{L}}(y)$ by $\max_{\ell\in\mathcal{L}}f_\ell(y)$ and the threshold by zero, which discards exactly the prior information that \eqref{eq:tau_bayes} shows to dominate the decision.
\end{remark}

\subsection{Neyman-Pearson Interpretation and Receiver Design Parameter}\label{subsec:np_reading}

The receiver-side design problem of Sec.~\ref{subsec:metrics} sits naturally in the \emph{Neyman-Pearson} framework~\cite{NP1933,Poor1994detection}: with idle ($X\in\mathcal{L}$) as $H_0$ and active ($X\in\mathcal{X}$) as $H_1$, the false-alarm probability $P_{\mathrm{fa}}$ is the Type-I and the erasure probability $P_{\mathrm{ers}}$ the Type-II error rate. The receiver-side design problem is the textbook Neyman-Pearson direction: minimize the Type-II rate subject to a hard cap on the Type-I side, here the silent-corruption rate \eqref{eq:Psilent} that carries $P_{\mathrm{fa}}$.

Fixing $\tau$ at $\tau_{\mathrm{Bayes}}$ would minimize the aggregate $P_{\mathrm{err}}$ of \eqref{eq:Perr_total}. The design problem of Sec.~\ref{subsec:metrics} instead caps the silent-corruption rate $P_{\mathrm{silent}}$ of \eqref{eq:Psilent} at $\bar P_{\mathrm{silent}}$ and minimizes the erasure rate, with all three metrics now read as functions $P_{\mathrm{con}}(\tau)$, $P_{\mathrm{ers}}(\tau)$, $P_{\mathrm{fa}}(\tau)$ of the threshold.

Under the rule \eqref{eq:decision}, the three metrics decouple as follows. The output-level confusion $P_{\mathrm{con}}(\tau)$ counts the joint event $\{\Lambda(Y)>\tau,\,\hat x(Y)\neq X\}$ given $X\in\mathcal{X}$, so it does depend on $\tau$; but it is upper-bounded for every $\tau$ by the pure codeword-\ac{ml} pair-confusion probability $\Pr\{\hat x(Y)\neq X\mid X\in\mathcal{X}\}$, which depends on codebook geometry and \ac{snr} alone. Picking a codebook whose pair-confusion contribution to \eqref{eq:Psilent} is negligible at the operating \ac{snr} is a codebook-selection step independent of $\tau$. The false-alarm probability $P_{\mathrm{fa}}(\tau)=\mathrm{Pr}\{\Lambda(Y)>\tau\mid X\in\mathcal{L}\}$ is monotonically decreasing in $\tau$, the erasure probability $P_{\mathrm{ers}}(\tau)=\mathrm{Pr}\{\Lambda(Y)\leq\tau\mid X\in\mathcal{X}\}$ is monotonically increasing, and the silent-corruption rate $P_{\mathrm{silent}}(\tau)$ is therefore monotonically decreasing.

The receiver-side design problem reduces, at fixed litter distribution $P_L$, to the one-dimensional program
\begin{equation}
\tau^*(P_L)\triangleq\arg\min_\tau P_{\mathrm{ers}}(\tau;P_L)\quad\text{s.t.}\quad P_{\mathrm{silent}}(\tau;P_L)\leq\bar P_{\mathrm{silent}},
\label{eq:tau_star}
\end{equation}
whose optimum is the smallest $\tau$ for which $P_{\mathrm{silent}}(\tau;P_L)\leq\bar P_{\mathrm{silent}}$, since $P_{\mathrm{silent}}$ is monotonically decreasing while $P_{\mathrm{ers}}$ is monotonically increasing. Sweeping $\tau$ traces a curve in the $(P_{\mathrm{fa}},P_{\mathrm{ers}})$ plane at fixed codebook, \ac{snr}, and $P_L$, along which the silent-corruption cap picks the operating point. The dependence of $\tau^*$ on $P_L$ marked in \eqref{eq:tau_star} couples the receiver back to the transmitter-side shaping, and is the inner subproblem of the joint mechanism design of Sec.~\ref{sec:shaping}.

\section{Threat Model and Indistinguishability Metric}\label{sec:threat}

The receiver-side analysis of Sec.~\ref{sec:receiver} treated the litter distribution $P_L$ as exogenous. The remaining design question is how to shape $P_L$ so that a third-party observer cannot reliably tell active from idle slots while the legitimate receiver retains its operational metrics. Two commitments make that question precise: an observer with stated knowledge and observation window, and a scalar measuring how far the idle output distribution stands from the active one at her receiver.

\subsection{Observer Model}\label{subsec:observer}

We model the adversary as a single passive observer, Eve, who receives every transmitted block over her own link. The two links have the form of Sec.~\ref{subsec:channel}, with gains $h^{\mathrm{Bob}}$ and $h^{\mathrm{Eve}}$ and the same noise floor $N_0$:
\begin{align}
Y^{\mathrm{Bob}} &= h^{\mathrm{Bob}}\mu(X) + Z^{\mathrm{Bob}},\quad Z^{\mathrm{Bob}}\sim\mathcal{CN}(0,N_0\,I_n),\label{eq:bob_channel}\\
Y^{\mathrm{Eve}} &= h^{\mathrm{Eve}}\mu(X) + Z^{\mathrm{Eve}},\quad Z^{\mathrm{Eve}}\sim\mathcal{CN}(0,N_0\,I_n).\label{eq:eve_channel}
\end{align}
Coherent scaling by the known gain brings each link to the form \eqref{eq:awgn} at per-symbol \ac{snr} $\gamma^{\mathrm{Bob}}=|h^{\mathrm{Bob}}|^2/N_0$ and $\gamma^{\mathrm{Eve}}=|h^{\mathrm{Eve}}|^2/N_0$, respectively, and every density below is evaluated at the corresponding \ac{snr}. As in any modern physical layer, we assume that the code sees enough diversity (in frequency, time, or space) for small-scale fading to be averaged out within a block, so that each gain is a large-scale quantity set by path loss and shadowing and varying slowly relative to the block. The same large-scale statistics govern both links: Eve's gain varies on the same slow time scale as Bob's, and the two links differ in what the transmitter knows about them, not in their physics. Every quantity in this paper is conditional on the two \acp{snr}.

The information asymmetry follows from where the pilots are measured, and is worth stating in full because the design turns on it. Every block, litter or codeword alike, opens with the same standard pilot preamble, so the preamble carries no activity information and each receiver estimates its own gain from it. Bob therefore measures $\gamma^{\mathrm{Bob}}$ and reports it, which is the \ac{csi} feedback that drives code and rate adaptation anyway, so both ends of the legitimate link hold the instantaneous $\gamma^{\mathrm{Bob}}$; we take the $(n,k)$ code as given at each operating point, evaluate every operational metric of Sec.~\ref{sec:receiver} at the reported value, and let the look-up table of Sec.~\ref{subsec:fading} follow its slow variation. Eve reads the same preamble and estimates $\gamma^{\mathrm{Eve}}$ from it just as accurately, so her tests in Sec.~\ref{subsec:hypotheses} run at her true \ac{snr}. Neither party observes the other's gain block by block. Eve's distance, antenna, and noise floor lie outside link-protocol control and no feedback from Eve is assumed, so the transmitter knows her link only statistically, through a prior $\mu_{\mathrm{Eve}}$ on $\gamma^{\mathrm{Eve}}$ centered below $\gamma^{\mathrm{Bob}}$; the analysis holds for any such prior, including a point mass for a transmitter that knows a single nominal value, and Sec.~\ref{sec:numerics} specifies the prior used numerically. Symmetrically, Eve is not assumed to read Bob's \ac{csi} feedback, so she knows the legitimate link only through its large-scale statistics. Table~\ref{tab:knowledge} collects this.

Bob's gain matters to Eve for one reason only: it selects the look-up-table row, hence the litter distribution $P_L$ she must test against. Rather than let her infer the row, we grant it to her, as part of the public-design assumption below. An observer who knows $P_L$ exactly is strictly stronger than one who must average over the prior on $\gamma^{\mathrm{Bob}}$, so every guarantee reported here holds a fortiori for the weaker observer that the pilot geometry alone would produce. The design exploits a gap in link quality, not in knowledge: Bob decodes reliably on his better channel, and Eve, who knows her own channel and the design in force, separates idle from active less reliably on her typically worse one. The gap is a matter of degree rather than of ordering. The two links shadow independently, so a $6$~dB median advantage is overturned whenever the shadowing difference exceeds it, which for $\sigma=6$~dB happens in about one realization in six; $\bar D$ averages over those realizations as well. Ruling them out would assume a deployment geometry in which the observer is never the better placed of the two, which the fading model does not supply.

\begin{table}[t]
\centering
\caption{Knowledge assumed of each party. \emph{Exact} denotes the instantaneous value, \emph{statistical} the large-scale distribution only.}
\label{tab:knowledge}
\begin{tabular}{l|ccc}
\toprule
 & Alice & Bob & Eve\\
\midrule
Codebook $\mathcal{X}$, activity rate $p$ & yes & yes & yes\\
Litter distribution $P_L$ in force & yes & yes & granted\\
Legitimate-link \ac{snr} $\gamma^{\mathrm{Bob}}$ & exact & exact & statistical\\
Observer-link \ac{snr} $\gamma^{\mathrm{Eve}}$ & statistical & statistical & exact\\
Activity state $A_t$ & yes & inferred & inferred\\
\bottomrule
\end{tabular}
\end{table}

Eve is otherwise modeled with three standard attributes. \textbf{Public design:} Eve knows the codebook $\mathcal{X}$, the activity prior $p$, and the litter distribution $P_L$ in use, the last as granted above; there is no shared secret between transmitter and Bob to which Eve is denied access. \textbf{Multi-block observation:} Eve observes a window of $T$ blocks $Y^{\mathrm{Eve},T}=(Y^{\mathrm{Eve}}_1,\dots,Y^{\mathrm{Eve}}_T)$ and may process it in any way. \textbf{Passive:} Eve does not jam or otherwise interfere with the channel; her only action is inference from $Y^{\mathrm{Eve},T}$.

\subsection{Eve's Hypothesis Tests}\label{subsec:hypotheses}

Write the density \eqref{eq:awgn_density} with explicit \ac{snr} parameter,
\begin{equation}
f_s(y;\gamma)\triangleq(\gamma/\pi)^{n}\exp\!\bigl(-\gamma\|y-\mu(s)\|^2\bigr),\quad s\in\mathcal{M}^n,\,\gamma>0.
\label{eq:awgn_density_param}
\end{equation}
Let $A_t\in\{\mathrm{act},\mathrm{idle}\}$ denote the activity state of block $t$. Conditioned on $A_t$ and on $\gamma^{\mathrm{Eve}}$, Eve's observation $Y^{\mathrm{Eve}}_t$ is drawn from the active or idle output mixture
\begin{align}
P_{Y\mid\mathrm{act}}^{\mathrm{Eve}}(y;\gamma^{\mathrm{Eve}}) &\triangleq M^{-k}\sum_{x\in\mathcal{X}}f_x(y;\gamma^{\mathrm{Eve}}),\label{eq:Py_active}\\
P_{Y\mid\mathrm{idle}}^{\mathrm{Eve}}(y;\gamma^{\mathrm{Eve}}) &\triangleq \sum_{\ell\in\mathcal{L}}P_L(\ell)\,f_\ell(y;\gamma^{\mathrm{Eve}}).\label{eq:Py_idle}
\end{align}
and unconditionally from the mixture $P_{Y}^{\mathrm{Eve}}(y;\gamma^{\mathrm{Eve}})\triangleq p\,P_{Y\mid\mathrm{act}}^{\mathrm{Eve}}(y;\gamma^{\mathrm{Eve}})+(1-p)\,P_{Y\mid\mathrm{idle}}^{\mathrm{Eve}}(y;\gamma^{\mathrm{Eve}})$. Two questions are open to Eve, and both are answered from the per-block likelihood ratio of \eqref{eq:Py_active} to \eqref{eq:Py_idle}. Whether a given block is active: the states $A_t$ are i.i.d. and the blocks are conditionally independent given $\gamma^{\mathrm{Eve}}$, so the posterior of $A_t$ given the whole window $Y^{\mathrm{Eve},T}$ equals its posterior given $Y^{\mathrm{Eve}}_t$ alone, and no processing of the sequence improves on a block-by-block likelihood-ratio test. Whether the transmitter has been active at all over the window: this is the traffic-analysis question of Sec.~\ref{sec:intro}, a test of the all-idle product distribution $(P_{Y\mid\mathrm{idle}}^{\mathrm{Eve}})^{\otimes T}$ against the mixture product distribution $(P_{Y}^{\mathrm{Eve}})^{\otimes T}$, both at Eve's realized $\gamma^{\mathrm{Eve}}$, which is constant over the window.

\subsection{Indistinguishability Metric}\label{subsec:KL_metric}

For a given realization of $\gamma^{\mathrm{Eve}}$, the per-block \ac{kl} divergence
\begin{equation}
D(P_L;\gamma^{\mathrm{Eve}}) \triangleq \int P_{Y\mid\mathrm{idle}}^{\mathrm{Eve}}(y;\gamma^{\mathrm{Eve}})\log\frac{P_{Y\mid\mathrm{idle}}^{\mathrm{Eve}}(y;\gamma^{\mathrm{Eve}})}{P_{Y\mid\mathrm{act}}^{\mathrm{Eve}}(y;\gamma^{\mathrm{Eve}})}\,\diff y
\label{eq:KL_per_realization}
\end{equation}
governs both of Eve's tests. Up to sign, it is the mean under idle of the per-block log-likelihood ratio that the block-by-block test thresholds, and by convexity of the divergence in its second argument the exponent of the presence test obeys $D\bigl(P_{Y\mid\mathrm{idle}}^{\mathrm{Eve}}\,\|\,P_{Y}^{\mathrm{Eve}}\bigr)\le p\,D(P_L;\gamma^{\mathrm{Eve}})$, so that by Stein's lemma~\cite{Poor1994detection} Eve needs at least $T\approx\log(1/\eta)/\bigl(p\,D(P_L;\gamma^{\mathrm{Eve}})\bigr)$ blocks to reach miss probability $\eta$ at a fixed false-alarm level. We take \eqref{eq:KL_per_realization} as the indistinguishability measure because it depends on $P_L$ alone and bounds Eve's exponent for every activity rate. Since the transmitter does not know Eve's realization, the design objective is its expectation over the prior $\mu_{\mathrm{Eve}}$ of Sec.~\ref{subsec:observer},
\begin{equation}
\bar D(P_L) \triangleq \mathbb{E}_{\gamma^{\mathrm{Eve}}\sim\mu_{\mathrm{Eve}}}\bigl[D(P_L;\gamma^{\mathrm{Eve}})\bigr],
\label{eq:KL_metric}
\end{equation}
left generic throughout the analysis. Two remarks fix what \eqref{eq:KL_metric} promises. It is the expected exponent, not the exponent of Eve's expected error: by Jensen's inequality $\mathbb{E}[e^{-TD}]\ge e^{-T\bar D}$, so $e^{-T\bar D}$ lower-bounds Eve's average miss probability, and for long windows that average is dominated by the weakest realization of $\gamma^{\mathrm{Eve}}$. A guarantee against every realization instead needs the largest exponent on the support of $\mu_{\mathrm{Eve}}$, which exists only if that support is bounded. Log-normal shadowing does not bound it, and bounding it by fiat would assert that the observer is never the better placed of the two, an assumption about deployment geometry rather than about fading: at the $6$~dB median gap and $\sigma=6$~dB of Sec.~\ref{subsec:observer}, independent shadowing overturns the gap in about one realization in six. We therefore report the expected exponent and say what it does not cover. Where a deployment does constrain the observer's position, so that $\gamma^{\mathrm{Eve}}\le\gamma^{\mathrm{Eve}}_{\max}$, the worst case follows from the same estimator as the point-mass prior at $\gamma^{\mathrm{Eve}}_{\max}$, since $D(P_L;\gamma)$ increased with $\gamma$ in all our evaluations. Sec.~\ref{sec:numerics} takes the expectation over the full prior instead. The transmitter shapes $P_L$ to drive $\bar D(P_L)$ as small as possible at each value of $\gamma^{\mathrm{Bob}}$, exploiting the known $\gamma^{\mathrm{Bob}}$ to deliver legitimate-link reliability while Eve's typically lower \ac{snr} limits her distinguishability between idle and active mixtures.

Convexity carries through. At each fixed $\gamma^{\mathrm{Eve}}$, $D(P_L;\gamma^{\mathrm{Eve}})$ is convex in $P_L$ on the simplex $\Delta(\mathcal{L})$: $P_{Y\mid\mathrm{idle}}^{\mathrm{Eve}}$ is linear in $P_L$ by \eqref{eq:Py_idle}, $P_{Y\mid\mathrm{act}}^{\mathrm{Eve}}$ is fixed by the public codebook and Eve's \ac{snr}, and \ac{kl} is convex in its first argument under a fixed reference. Expectation over $\mu_{\mathrm{Eve}}$ preserves convexity. The frozen-statistic shaping subproblem of Sec.~\ref{subsec:shaping_relax} therefore retains its convex structure when $D(P_L)$ is replaced by $\bar D(P_L)$; the \ac{map}-aware case treated by the alternating algorithm of Sec.~\ref{subsec:shaping_relax} and the \ac{drl} solver of Sec.~\ref{subsec:shaping_drl} introduces a $P_L$-dependent receiver threshold that breaks the joint program's convexity even though $\bar D$ itself is well-behaved.\footnote{The metric \eqref{eq:KL_metric} is distinct from asymptotic channel resolvability~\cite{HV1993approximation,Hayashi2006general}, which asks how much input randomness drives a channel output toward a target distribution and supplies the constructive tool behind covertness proofs~\cite{Bloch2016covert}; $\bar D$ asks how distinguishable two given output mixtures are at the observer, as a per-block design objective.}


\subsection{CSI-Feedback Adaptation Architecture}\label{subsec:fading}

Bob's gain $h^{\mathrm{Bob}}$, hence $\gamma^{\mathrm{Bob}}$, changes on the slow time scale of large-scale channel variation and is tracked by the \ac{csi} feedback, so the design is a function of $\gamma^{\mathrm{Bob}}$ and is implemented as a look-up table. Offline, the Stackelberg game \eqref{eq:stackelberg} of Sec.~\ref{sec:shaping} is solved on a grid $\gamma_1<\dots<\gamma_J$ of Bob's \ac{snr}, and row $j$ of the look-up table stores the class-level litter distribution $P_C^*(\gamma_j)$ of Sec.~\ref{subsec:shaping_practical} together with the threshold $\tau^*(\gamma_j)$: $J(C+1)$ numbers, a few kilobytes at the $(16,12)$ geometry. Online, both ends index the look-up table at every block boundary by the most recent \ac{csi} report, quantized down to the grid, in the same way as a rate-adaptation table. The transmitter fills each idle slot by drawing a class from $P_C^*(\gamma_j)$ and then a uniform member of that class, one encoder pass for the coset classes of Sec.~\ref{subsec:shaping_practical}; the receiver forms the litter aggregate \eqref{eq:LL} from the same $P_C^*(\gamma_j)$ and applies $\tau^*(\gamma_j)$, optionally refined online through \eqref{eq:tau_star}. No optimization runs online. Eve's side needs no adaptation, since her gain enters only through the prior $\mu_{\mathrm{Eve}}$, and public design means that she holds the same look-up table. The caps of Sec.~\ref{subsec:metrics} hold on every grid point by construction; between two grid points the entry of the lower one is used, which is conservative because all three error rates of Sec.~\ref{subsec:metrics} fall with Bob's \ac{snr} at a fixed design.

The look-up table is built at a fixed $(n,k)$. A deployed link adapts its modulation and coding scheme to $\gamma^{\mathrm{Bob}}$ as well, in which case the same \ac{csi} index selects a scheme and the look-up table carries one row per (\ac{snr}, scheme) pair; the construction is unchanged, and choosing the scheme is the ordinary link-adaptation problem, settled offline before the cover design is laid on top of it. The numerical study of Sec.~\ref{sec:numerics} holds the scheme fixed and varies only the \ac{snr}.
\section{Optimal Litter Distribution Shaping}\label{sec:shaping}

Sec.~\ref{sec:receiver} optimized the threshold $\tau$ at a fixed litter distribution, and Sec.~\ref{sec:threat} fixed the observer that the distribution has to defeat. Letting $P_L$ vary closes the loop between the two, with the receiver's threshold response \eqref{eq:tau_star} entering as an inner subproblem. What results is a Stackelberg game: the transmitter, as leader, picks the litter distribution that minimizes the observer's distinguishing power, and the receiver, as follower, best-responds with the threshold $\tau^*(P_L)$ that minimizes the erasure rate under the silent-corruption cap.

\subsection{Stackelberg Formulation}\label{subsec:shaping_stackelberg}

Let $\Delta(\mathcal{L})\triangleq\{P_L\in\mathbb{R}^{|\mathcal{L}|}_{\ge 0}:\sum_{\ell\in\mathcal{L}}P_L(\ell)=1\}$ denote the probability simplex on $\mathcal{L}$. The Stackelberg game writes as, for each value of $\gamma^{\mathrm{Bob}}$,
\begin{subequations}\label{eq:stackelberg}
\begin{align}
\min_{P_L\in\Delta(\mathcal{L})} \quad & \bar D(P_L)\label{eq:stackelberg_obj}\\
\mathrm{s.t.}\quad & P_{\mathrm{ers}}\bigl(\tau^*(P_L);P_L\bigr)\le\bar P_{\mathrm{ers}},\label{eq:stackelberg_erscap}\\
& \tau^*(P_L)\ \text{solves the follower program}~\eqref{eq:tau_star}.\label{eq:stackelberg_follower}
\end{align}
\end{subequations}
The outer problem~\eqref{eq:stackelberg_obj} is the indistinguishability objective the transmitter minimizes against the public-design observer of Sec.~\ref{sec:threat}, taking the expectation over the prior $\mu_{\mathrm{Eve}}$ on Eve's \ac{snr}. The inner equilibrium condition~\eqref{eq:stackelberg_follower} is the receiver's best response at the current \ac{csi}-reported $\gamma^{\mathrm{Bob}}$, identical to the receiver-side design problem of Sec.~\ref{subsec:np_reading} now seen as a $P_L$-parameterized map; the silent-corruption cap $\bar P_{\mathrm{silent}}$ lives inside it and is therefore enforced at every choice of $P_L$. The outer feasibility constraint~\eqref{eq:stackelberg_erscap} is the throughput condition: the erasure rate the receiver is left with after meeting that cap must stay within the retransmission budget $\bar P_{\mathrm{ers}}$, since every erased block is a block the upper layer must send again. The look-up table of Sec.~\ref{subsec:fading} collects the solutions $\bigl(P_L^*(\gamma^{\mathrm{Bob}}),\tau^*(\gamma^{\mathrm{Bob}})\bigr)$ of \eqref{eq:stackelberg} across the grid of $\gamma^{\mathrm{Bob}}$.

\begin{remark}[The two caps commute]\label{rem:commute}
Interchanging the roles of the two caps, so that the follower minimizes $P_{\mathrm{silent}}$ subject to $P_{\mathrm{ers}}\le\bar P_{\mathrm{ers}}$ and the leader carries the silent-corruption cap as its outer constraint, leaves the problem in $P_L$ unchanged. Both orders declare $P_L$ feasible exactly when some threshold meets both caps at once: with $P_{\mathrm{silent}}$ decreasing and $P_{\mathrm{ers}}$ increasing in $\tau$, the smallest $\tau$ meeting the silent cap minimizes erasure among all thresholds that meet it, and the largest $\tau$ meeting the erasure cap minimizes silent corruption among all thresholds that meet that one. Since $\bar D(P_L)$ does not depend on $\tau$, the feasible set and the optimizer $P_L^*$ coincide; only the threshold at which the design is operated differs, and with it the erasure the link actually pays. We solve \eqref{eq:stackelberg} in the erasure-capped order, which keeps the coefficients of the inner program of Sec.~\ref{subsec:shaping_relax} affine, and report both operating points in Sec.~\ref{sec:numerics}: the \emph{design point} at the erasure cap and the \emph{deployed point} at the silent cap.
\end{remark}

The joint program \eqref{eq:stackelberg} is non-convex because the \ac{map} test statistic $\Lambda(y;P_L)=\log[L_{\mathcal{X}}(y)/L_{\mathcal{L}}(y;P_L)]$ depends on $P_L$: the constraint is not affine in $P_L$, and the implicit $\tau^*(P_L)$ adds further coupling. The outer objective $\bar D(P_L)$ remains convex throughout, as shown in Sec.~\ref{subsec:KL_metric}, but the feasible set is not. Sec.~\ref{subsec:shaping_relax} develops a convex-relaxation route as a structured baseline; Sec.~\ref{subsec:shaping_drl} presents a \ac{drl} solver that attacks the coupling directly, without a convex surrogate. Neither method carries a global-optimum guarantee, which is why we run both: they descend the same landscape by unrelated routes, so their agreement is evidence that neither is stranded far from the achievable optimum. We take the minimizer across them as the deployed $P_L^*$.

\subsection{Decomposition and Alternating Convex Relaxation}\label{subsec:shaping_relax}

At fixed $P_L$, the receiver's best response is found by one-dimensional binary search on a monotone curve, since $P_{\mathrm{silent}}(\tau;P_L)$ is monotonically decreasing while $P_{\mathrm{ers}}(\tau;P_L)$ is monotonically increasing, as Sec.~\ref{subsec:np_reading} established. Following Remark~\ref{rem:commute} we run the search on the erasure curve $P_{\mathrm{ers}}(\cdot;P_L)$ to locate the design threshold, which is what keeps the inner constraint below affine in $P_L$, and recover the deployed threshold by the matching search on $P_{\mathrm{silent}}(\cdot;P_L)$ once $P_L^*$ is fixed. The best response must be recomputed at every outer iterate because the test statistic $\Lambda(\cdot;P_L)$ reshapes with $P_L$.

Freezing the test statistic leaves a subproblem that is convex in $P_L$. At a reference $P_L^{(0)}\in\Delta(\mathcal{L})$, uniform by default, the receiver is held at the rule
\begin{equation}
\Lambda^{(0)}(y)\triangleq\log\frac{L_{\mathcal{X}}(y)}{L_{\mathcal{L}}(y;P_L^{(0)})},\qquad\tau^{(0)}=\tau^*(P_L^{(0)}).
\label{eq:Lambda_frozen}
\end{equation}
Under this rule the law of total probability gives $P_{\mathrm{fa}}^{(0)}(P_L)=\sum_\ell P_L(\ell)\,a_\ell^{(0)}$ with $a_\ell^{(0)}\triangleq\Pr\{\Lambda^{(0)}(Y)>\tau^{(0)}\mid X=\ell\}$ linear in $P_L$, and $P_{\mathrm{ers}}^{(0)},P_{\mathrm{con}}^{(0)}$ constant in $P_L$. The shaping subproblem at frozen reference then reads
\begin{equation}
\min_{P_L\in\Delta(\mathcal{L})}\bar D(P_L)\quad\mathrm{s.t.}\quad p\,P_{\mathrm{con}}^{(0)}+(1-p)\,P_{\mathrm{fa}}^{(0)}(P_L)\le\bar P_{\mathrm{silent}},
\label{eq:shaping_convex}
\end{equation}
a convex program on the simplex with a single affine constraint. \eqref{eq:shaping_convex} relaxes \eqref{eq:stackelberg} by replacing the \ac{map}-aware threshold response with the frozen reference; the erasure cap is carried by the frozen threshold $\tau^{(0)}$, which sits at that cap by construction.

Algorithm~\ref{alg:alt} refines the frozen-statistic relaxation by iterating: update the reference to the previous iterate, recompute $\tau^*$, and re-solve \eqref{eq:shaping_convex}. Each re-solve is confined to a chi-square trust region around its reference, which bounds the variance of the importance weights that re-anchor the sample-average approximation of $\bar D$ between reference updates; the experiments of Sec.~\ref{sec:numerics} cap the chi-square distance at $1$, which guarantees an effective sample fraction of at least one half. The iterates decreased monotonically in $\bar D$ throughout our experiments; we claim no stationarity or global-optimum guarantee on \eqref{eq:stackelberg}. The converged $P_L^{(k)}$ is feasible for \eqref{eq:stackelberg} by construction and gives a documented upper bound on its optimal value.

One implementation detail of the sample-average step matters more than it looks. The importance sample at each outer iteration is drawn from the current reference, and every term of \eqref{eq:shaping_convex} is divided by the reference density at the sampled point. At high \ac{snr} the optimal design is sparse: the shaper drives most classes to exactly zero from the first iteration on. A sample drawn from that collapsed reference then sees densities near zero at the surviving low-weight classes, the divisor grows without bound, and the conic solve eventually fails; on the instance of Sec.~\ref{sec:numerics} the largest divisor rose from $255$ at the uniform reference to $8.5\times10^{6}$ by the fifth iteration, at which point the solver raised. The remedy is the standard one, a defensive mixture: the sample is drawn from, and the density evaluated at, $(1-\alpha)P_C^{(k)}+\alpha\,u$ with $u$ uniform on the classes and $\alpha=0.05$. Since each row of the class-likelihood matrix is normalised to a maximum of one, the mixture density is at least $\alpha\min_c u(c)$, which caps the divisor at $C/\alpha$, about $5\,100$ here, whatever the shaper does to the reference. The estimator remains unbiased for the distribution it now samples, because the same mixture supplies both the draws and the density. With the mixture the same sequence completes every iteration.

\begin{algorithm}[t]
\caption{Alternating convex relaxation for \eqref{eq:stackelberg}}
\label{alg:alt}
\KwIn{$\mathcal{X},\gamma^{\mathrm{Bob}},\mu_{\mathrm{Eve}},\bar P_{\mathrm{silent}},\bar P_{\mathrm{ers}},\delta$}
$P_L^{(0)}\gets$ uniform on $\mathcal{L}$;\ \ $k\gets 0$\;
\Repeat{$|\bar D(P_L^{(k)})-\bar D(P_L^{(k-1)})|<\delta$}{
  $\tau^{(k)}\gets\tau^*(P_L^{(k)})$ by binary search on $P_{\mathrm{ers}}(\cdot;P_L^{(k)})$\;
  Compute coefficients $a_\ell^{(k)}$ at $(\Lambda(\cdot;P_L^{(k)}),\tau^{(k)})$\;
  Solve \eqref{eq:shaping_convex} at reference $P_L^{(k)}$, within the chi-square trust region, for $P_L^{(k+1)}$;\quad$k\gets k+1$\;
}
$\tau_{\mathrm{dep}}\gets$ smallest $\tau$ with $P_{\mathrm{silent}}(\tau;P_L^{(k)})\le\bar P_{\mathrm{silent}}$\;
\Return $P_L^{(k)}$ with design threshold $\tau^{(k)}$ and deployed threshold $\tau_{\mathrm{dep}}$\;
\end{algorithm}

\subsection{Deep-Reinforcement-Learning Solver}\label{subsec:shaping_drl}

We solve \eqref{eq:stackelberg} directly by a \ac{drl} policy that maps a problem instance to a litter distribution, trained offline on simulated channel and observer realizations. The \ac{drl} solver meets the \ac{map}-statistic coupling head on, through adaptive constraint penalization and with no convex-program reformulation, so it neither freezes the test statistic nor confines its steps to a trust region. It is therefore an independent check on the alternating relaxation rather than a refinement of it: the two share no approximation, so where they agree, the agreement is informative.

Among reinforcement-learning families, on-policy clipped policy gradient, \ac{ppo}~\cite{SWD+2017ppo}, is the canonical estimator for one-step contextual bandits at moderate action dimension, and it supports the stochastic simplex-valued policies our action space requires. The alternatives each lose to the problem structure: replay-based off-policy actor-critics~\cite{HZAL2018sac} cannot exploit the one-step formulation, deterministic actors~\cite{LHP+2016ddpg} cannot represent mixed strategies on $\Delta_C$, and Stackelberg-aware hypergradient corrections~\cite{ZFA+2022stackac} collapse to ordinary backpropagation once the follower's best response is the closed-form $\tau^*(P_L)$ of Sec.~\ref{subsec:shaping_relax}. We therefore adopt \ac{ppo}, with the three adaptations that the litter-shaping geometry dictates.

We cast \eqref{eq:stackelberg} as a contextual one-step \ac{mdp}. The \emph{context} $s=(\gamma^{\mathrm{Bob}},\mu_{\mathrm{Eve}},\mathcal{X})$ summarizes the operating point. The \emph{action} is a class-restricted distribution $P_C\in\Delta_C$ over the $C$ distance-equivalence classes of $\mathcal{L}$ introduced in Sec.~\ref{subsec:shaping_practical}. The \emph{reward} embeds the constraint as a one-sided penalty,
\begin{equation}
r(s,P_C)=-\bar D(P_L)-\lambda_{t}\bigl[P_{\mathrm{silent}}(\tau^*(P_L);P_L)-\bar P_{\mathrm{silent}}\bigr]_+,
\label{eq:drl_reward}
\end{equation}
where $P_L$ is the class-restricted expansion of $P_C$, $[x]_+\triangleq\max(0,x)$, and $\lambda_{t}\ge 0$ is a time-varying Lagrange multiplier updated by \eqref{eq:pid_lambda}. The threshold $\tau^*(P_L)$ sits at the erasure cap, so the penalized quantity is the silent-corruption rate there, which is the erasure-capped order of Remark~\ref{rem:commute}. The receiver's threshold $\tau^*(P_L)$ is computed by the binary search of Sec.~\ref{subsec:shaping_relax} during reward evaluation and is not learned.

The action $P_C\in\Delta_C$ is a probability distribution on the class simplex, and this is the first place where a standard solver needs adapting: the default \ac{ppo} Gaussian-softmax head is biased on $\Delta_C$ at moderate-to-large $C$~\cite{TKB+2022dirichlet}. We replace it with a Dirichlet head: the policy network outputs concentration logits $z=\mathrm{NN}_\theta(s)\in\mathbb{R}^C$, the Dirichlet parameters are $\boldsymbol\alpha=\mathrm{softplus}(z)+10^{-3}\mathbf{1}$, and $P_C\sim\mathrm{Dir}(\boldsymbol\alpha)$. The Dirichlet head is bias-free on the simplex and admits closed-form log-likelihood and entropy for the \ac{ppo} ratio. The policy body is two hidden layers of $64$ ReLU units; advantages are computed against an exponential-moving-average reward baseline, which the one-step structure makes sufficient in place of a learned value function.

The penalty weight in \eqref{eq:drl_reward} is set each iteration by the \ac{pid}-Lagrangian controller of Stooke \emph{et~al.}~\cite{SAA2020pidlagrangian},
\begin{equation}
\begin{aligned}
\lambda_{t}&=\bigl[K_P\,\bar g_{t}+I_{t}+K_D\,(\bar g_{t}-\bar g_{t-1})\bigr]_+,\\
I_{t}&=\bigl[I_{t-1}+K_I\,\bar g_{t}\bigr]_+,
\end{aligned}
\label{eq:pid_lambda}
\end{equation}
where $\bar g_{t}\ge 0$ is the rollout-averaged constraint violation at iteration $t$ and the integrator $I_{t}$ is clipped at zero so that satisfied periods bleed the multiplier off without driving it negative. The proportional and derivative terms damp the oscillations of a pure dual-ascent update, sparing the per-operating-point retuning that frozen penalty weights require.

Rather than parameterizing the policy over the full litter simplex $\Delta(\mathcal{L})$ at $|\mathcal{L}|\sim 10^9$, we restrict the action to the class simplex $\Delta_C$ at $C\sim 10^2$-$10^3$. The reward evaluation expands $P_C$ to the litter-level $P_L$ for the \ac{kl} and constraint computations through the closed-form linear class-to-litter map; the policy gradient flows back to $P_C$ through the same map. Without this restriction the policy network's output dimension is intractable; with it, the network is moderate-size and trainable on a single \ac{gpu}.

Each \ac{ppo} iteration draws $N$ candidate distributions from the current policy, evaluates each reward with $K$ Eve-\ac{snr} draws and Monte-Carlo output samples at every draw, updates the policy by a clipped policy-gradient step against the moving-average baseline, and updates the Lagrange multipliers by \eqref{eq:pid_lambda}; one iteration therefore costs $N$ reward evaluations plus one gradient pass.


In deployment the trained policy is queried for the current context $s_{\mathrm{deploy}}$ and outputs $P_C^{\mathrm{deploy}}=\boldsymbol\alpha/\sum_c\alpha_c$ (the Dirichlet mean) in a single forward pass, with the paired threshold taken from the same look-up entry and optionally refined via \eqref{eq:tau_star}; both run in well under a millisecond on commodity hardware.

The policy is a two-layer perceptron with $64$ units per layer feeding the Dirichlet concentration head; \ac{ppo} uses clip ratio $\epsilon=0.2$, learning rate $10^{-2}$, $N=32$ rollouts per iteration with $K=8$ Eve-\ac{snr} draws per reward evaluation, four epochs of minibatch size $8$ per iteration, and $T_{\mathrm{train}}=60$ iterations; \ac{pid}-Lagrangian gains are $K_P=1$, $K_I=10^{-2}$, $K_D=10^{-1}$. The policy warm-starts at the alternating iterate, $\boldsymbol\alpha=200\,P_C^{\mathrm{alt}}$, refining a feasible design rather than re-exploring from uniform. Seeds, software versions, and configuration files accompany the released code. Convergence is to a local stationary point, not the global optimum of \eqref{eq:stackelberg}.

\subsection{Practical Computation}\label{subsec:shaping_practical}

Two reductions make the framework tractable. \emph{Class restriction}: the simplex $\Delta(\mathcal{L})$ has dimension $|\mathcal{L}|=M^n-M^k$, which is prohibitive for direct optimization at moderate $n$. We therefore partition $\mathcal{L}$ into equivalence classes and optimize a class-level distribution $P_C$ over the $C\ll|\mathcal{L}|$ classes, with $P_L$ taken uniform within each class; the per-litter false-alarm coefficients aggregate into per-class coefficients, and \eqref{eq:shaping_convex} reduces to a convex program on the $C$-dimensional simplex. The partition itself is rate-adaptive. At low rate we declare $\ell,\ell'\in\mathcal{L}$ equivalent if their Euclidean-distance profiles to $\mathcal{X}$ coincide, estimating the classes by sampling, which yields $C$ of order $10^2$ at rate $1/2$ and short blocklength. At high rate, including the $(16,12)$ geometry of Sec.~\ref{sec:numerics}, we instead use the cosets of the binary inner code: at the bit level, $\mathcal{L}$ is exactly the union of the $M^{n-k}-1$ nonzero cosets of the code, so class membership is a single syndrome computation, every class has exactly $M^k$ elements, and neither the partition nor its class sizes requires enumerating $\mathcal{X}$. Coset classes are geometry-homogeneous: Gray-\ac{qpsk} makes squared Euclidean distance proportional to bit-level Hamming distance, and codeword translation permutes the code, so coset members share their distance profile to $\mathcal{X}$. With $C=M^{n-k}-1=255$ at $(16,12)$, each iteration of Algorithm~\ref{alg:alt} solves in seconds. \emph{Sample-average approximation}: the expectation in $\bar D(P_L)$ over $\mu_{\mathrm{Eve}}$ and the integral over $\mathbb{C}^n$ are replaced by an average over $K$ draws of $\gamma^{\mathrm{Eve}}\sim\mu_{\mathrm{Eve}}$ and, at each draw, an empirical average over output samples from the idle mixture; the resulting bias is controlled by $K$ and the per-draw sample count. In the high-rate regime the active-side mixture likelihood entering each sample is itself evaluated exactly in $O(M^{n-k})$ per observation by Poisson summation over the dual code, rather than by the $M^k$-term primal sum. Class restriction is essential: without it both Algorithm~\ref{alg:alt} and the \ac{drl} solver of Sec.~\ref{subsec:shaping_drl} are intractable at $|\mathcal{L}|\sim 10^9$ on codes of practical interest.

\section{Numerical Results}\label{sec:numerics}

We evaluate the shaping design of Sec.~\ref{sec:shaping} on a rate-$3/4$ $(16,12)$ \ac{ldpc} instance, swept over Bob's \ac{snr} and the activity rate. Every result carries the two operating points that Remark~\ref{rem:commute} distinguishes: the \emph{design point}, at which the threshold sits at the erasure cap and the shaper spends its silent-corruption budget, and the \emph{deployed point}, at which the threshold sits at the silent-corruption cap and the link pays the resulting erasure.

\subsection{Experimental Setup}\label{subsec:numerics_setup}

The transmitter uses a regular column-weight-3 \ac{ldpc} code with $n=16$ \ac{qpsk} symbols and $k=12$ information symbols (rate $3/4$), so the codebook has $|\mathcal{X}|=4^{12}\approx 1.68\times 10^{7}$ members and the litter set $|\mathcal{L}|=4^{16}-4^{12}\approx 4.28\times 10^{9}$, which the coset construction of Sec.~\ref{subsec:shaping_practical} partitions into $C=4^{4}-1=255$ classes. The receiver runs the belief-propagation inner decoder. Bob's \ac{snr} is known and is swept over $\{0,2,\dots,20\}$~dB, with activity rate $p\in\{0.01,0.1,0.5\}$ and five independent seeds per cell, for $165$ independent runs in all. Eve's \ac{snr} is drawn from a log-normal shadowing prior, standard deviation $6$~dB, with median $6$~dB below Bob's, matching the slow large-scale statistics both links share. Operational caps are $\bar P_{\mathrm{silent}}=10^{-3}$ and $\bar P_{\mathrm{ers}}=10^{-2}$.

The three shapers compared are: \textbf{(i)~uniform-on-litter}, the no-shaping baseline; \textbf{(ii)~alternating}, Algorithm~\ref{alg:alt} with a chi-square trust region $\chi^2\le 1$ per inner step and the importance sample drawn from the reference mixed with $5\%$ uniform; and \textbf{(iii)~\ac{drl}} with \ac{ppo} clipped policy gradient, Dirichlet output head, and \ac{pid}-Lagrangian multiplier, warm-started from the alternating solution. Each Monte-Carlo estimate uses $10^{5}$ block trials for the operational metrics and $1\,024$ Eve-\ac{snr} draws with $1\,024$ output samples per draw for $\bar D$. Both thresholds are found by quantile inversion on a separate budget of $2\times 10^{4}$ calibration trials, which resolves the $10^{-3}$ silent-corruption cap to about $20$ events and is the accuracy limit on the deployed point. All values report mean$\pm$standard error across the five seeds.

The sweep varies Bob's \ac{snr} at a fixed code, which measures how sensitive the cover design is to link quality at one coding rate. It is not a deployment profile. A real link adapts its modulation and coding scheme to the \ac{snr} and would never run a rate-$3/4$ code at $0$~dB; fixing the scheme is what isolates the effect of interest here, at the cost of a low-\ac{snr} region that link adaptation would never visit. Selecting the scheme per \ac{snr} is the ordinary link-adaptation study, run once offline by simulation to populate the look-up table of Sec.~\ref{subsec:fading}, and the cover design of this paper then applies within each entry.

\subsection{Indistinguishability Reduction}\label{subsec:numerics_barD}

\begin{figure}[t]
\centering
\includegraphics[width=0.80\linewidth]{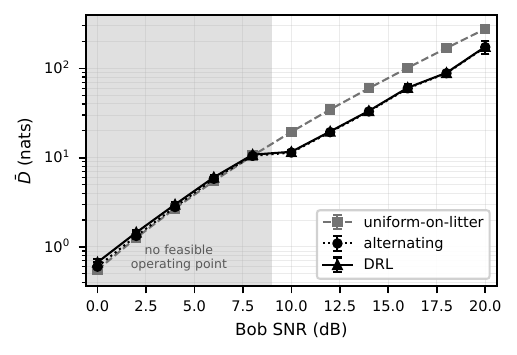}
\caption{Indistinguishability metric $\bar D$ versus Bob's \ac{snr} on the $(16,12)$ instance at $p=0.1$, mean and standard error over five seeds. No cover distribution meets both operational caps in the shaded region.}
\label{fig:bar_D_vs_snr}
\end{figure}

Fig.~\ref{fig:bar_D_vs_snr} plots $\bar D$ versus Bob's \ac{snr} at $p=0.1$, and Table~\ref{tab:bar_D} reports the values; the shaded region of the figure is the one Sec.~\ref{subsec:numerics_ops} finds inoperable, and the reduction column of the table refers to the better of the two shaped designs. Two regimes appear, and the boundary between them falls at $10$~dB.

Below $10$~dB shaping buys nothing, for two distinct reasons. Up to $4$~dB the inner program \eqref{eq:shaping_convex} is infeasible: codeword confusion alone exceeds the silent-corruption cap, the conic solver reports infeasibility in every one of the $45$ runs there, and the relaxation returns its uniform reference. The alternating curve at those points therefore \emph{is} uniform-on-litter, and the $4$--$9\%$ by which it appears worse is the disagreement between two Monte-Carlo estimates of one distribution. At $6$ and $8$~dB the program is feasible and the shaper moves, but the idle and active mixtures overlap so heavily that there is nothing to remove, and it lands within a few tenths of a nat of uniform. The \ac{drl} policy, whose penalty cannot be met where the program is infeasible, drifts $10$--$22\%$ above uniform at the lowest \acp{snr}; that is its exploration under an unsatisfiable constraint, not a property of shaped cover. None of this matters operationally, since Sec.~\ref{subsec:numerics_ops} shows no cover distribution admits an operating point below $10$~dB.

From $10$~dB upward the mixtures separate, the uniform baseline climbs faster than either shaped design, and the reduction settles onto a plateau of $41$--$48\%$ that neither grows nor decays with \ac{snr} and does not depend on the activity rate. This flatness is itself the finding: over a $10$~dB span in which $\bar D$ grows by a factor of $14$, shaping removes a fixed fraction of it. The gap is large relative to its uncertainty, between $6$ and $10$ standard errors at every point from $10$ to $18$~dB. The activity rate is immaterial to $\bar D$: it depends on the litter distribution alone, and $p$ enters only through the silent-corruption cap inside the follower program, so the measured values at $p\in\{0.01,0.5\}$ coincide with Fig.~\ref{fig:bar_D_vs_snr} within Monte-Carlo noise.

The two solvers land on the same answer. Across the $18$ cells from $10$~dB up the \ac{drl} policy is behind the alternating relaxation, by $0.3$ to $1.4$ percentage points of reduction, always within one standard error, and ahead of it in none. Since the two share no approximation, as Sec.~\ref{subsec:shaping_drl} notes, their agreement is the strongest evidence available here that both are close to what this problem admits. It also means the \ac{drl} solver earns its place as corroboration rather than as an improvement, and a practitioner may run the convex route alone.

One cell in Table~\ref{tab:bar_D} sits below the plateau, $20$~dB at $p=0.1$, at $38\%$ with a standard error four times its neighbours'. One of its five seeds ran all twelve iterations and finished $4\%$ above uniform; the other four sit at $43$--$52\%$. We report the cell as measured. Every one of the $120$ runs from $6$~dB upward completed all twelve iterations with the inner program solved at every step, so no cell in the table is an early stop.

\begin{table}[t]
\centering
\caption{Indistinguishability metric $\bar D$ (nats, mean and standard error over five seeds) on the $(16,12)$ instance at $p=0.1$, and its reduction relative to uniform cover.}
\label{tab:bar_D}
\begin{tabular}{c|ccc|c}
\toprule
$\gamma^{\mathrm{Bob}}$ (dB) & uniform-on-litter & alternating & \ac{drl} & reduction\\
\midrule
$6$  & $5.5\pm0.4$   & $5.8\pm0.4$   & $6.0\pm0.3$   & none\\
$8$  & $10.5\pm0.6$  & $10.5\pm0.6$  & $10.8\pm0.4$  & none\\
$10$ & $19.4\pm1.0$  & $11.4\pm0.5$  & $11.6\pm0.5$  & $41\%$\\
$12$ & $34.6\pm1.7$  & $19.3\pm0.9$  & $19.7\pm0.7$  & $44\%$\\
$14$ & $59.9\pm2.7$  & $32.9\pm1.5$  & $33.4\pm1.2$  & $45\%$\\
$16$ & $101\pm4$     & $59.7\pm5.2$  & $60.7\pm5.7$  & $41\%$\\
$18$ & $169\pm7$     & $88.2\pm3.0$  & $89.2\pm2.5$  & $48\%$\\
$20$ & $277\pm11$    & $173\pm29$    & $175\pm29$    & $38\%$\\
\bottomrule
\end{tabular}
\end{table}

\subsection{Operational Compliance}\label{subsec:numerics_ops}

At the design point the erasure rate sits at its cap $\bar P_{\mathrm{ers}}=10^{-2}$ for every method and every cell, to within the calibration noise of Sec.~\ref{subsec:numerics_setup}. What varies is the silent-corruption rate $P_{\mathrm{silent}}=pP_{\mathrm{con}}+(1-p)P_{\mathrm{fa}}$ left over there, and its two terms behave quite differently. The codeword-confusion rate $P_{\mathrm{con}}$ is fixed by codebook geometry, identical across $p$ and across shapers because no cover distribution can touch it; it falls from $0.97$ at $0$~dB through $4.4\times10^{-2}$ at $6$~dB and $1.1\times10^{-3}$ at $8$~dB to nothing measurable in $10^{5}$ trials at $12$~dB and above. The rate-$3/4$ code packs codewords densely, so they confuse readily at low \ac{snr}. The false-alarm rate $P_{\mathrm{fa}}$ is the term the shaper controls, and the shaper spends it: at $10$~dB and $p=0.1$ the design-point $P_{\mathrm{silent}}$ is $2.1\times10^{-4}$ under uniform cover but $1.1\times10^{-3}$ under either shaped design, which is the cap. The constraint is active, as it should be. Shaped litter is litter that sits closer to the codebook, and the silent-corruption budget is the currency it is bought with.

\begin{figure}[t]
\centering
\includegraphics[width=0.80\linewidth]{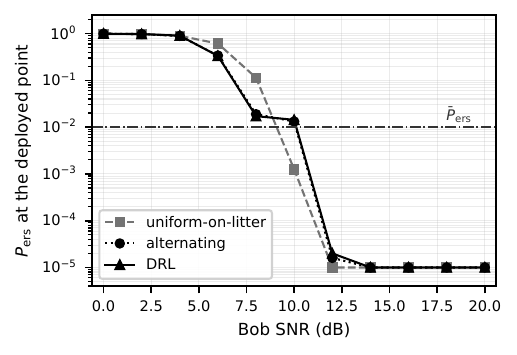}
\caption{Erasure rate at the deployed point, where the threshold holds the silent-corruption rate at $\bar P_{\mathrm{silent}}=10^{-3}$, versus Bob's \ac{snr} at $p=0.1$. Points on the floor are below the Monte-Carlo resolution.}
\label{fig:deployed_erasure}
\end{figure}

The price appears at the deployed point, where the threshold is moved to hold $P_{\mathrm{silent}}$ at $10^{-3}$ and the erasure rate is whatever remains. Fig.~\ref{fig:deployed_erasure} plots it at $p=0.1$; values at or below $10^{-5}$ are at the resolution of $10^{5}$ trials and are drawn on the floor of the figure. Four regimes follow, and together they set the operating envelope:
\begin{itemize}
\item From $0$ to $6$~dB, confusion alone exceeds the silent-corruption cap, so no threshold and no cover distribution can meet both caps. Holding the silent cap here erases between $99\%$ and $32\%$ of all transmitted blocks. The link is not operational at this rate, whatever the cover.
\item At $8$~dB there is still no feasible point, but the gap narrows and the cover distribution starts to matter: holding the silent cap costs uniform cover $11\%$ of blocks and shaped cover $1.7$--$1.9\%$, an improvement of roughly sixfold from shaping alone.
\item At $10$~dB uniform cover becomes feasible, at an erasure of $1.3\times10^{-3}$. The shaped designs sit at $1.3$--$1.4\times10^{-2}$, marginally above the cap and within the calibration noise of it. This is the one operating point where the trade is visible in both directions: shaping cuts $\bar D$ by $41\%$ and costs about one percent of throughput.
\item From $12$~dB upward, deployed erasure falls below $2\times10^{-5}$ for every method and is unmeasurable from $14$~dB. Here shaping is free: the full $40\%$ reduction in $\bar D$ costs no throughput at all.
\end{itemize}
The frontier below $10$~dB is a property of the code rather than of the cover, since it is set by $P_{\mathrm{con}}$; a lower-rate or stronger code would move it down at a cost in spectral efficiency. Above it, the regime of practical interest, the reductions of Fig.~\ref{fig:bar_D_vs_snr} are available, and from $12$~dB they are available for nothing.

\subsection{Discussion and Limitations}\label{subsec:numerics_limitations}

Four remarks put these results in context. First, the feasibility frontier of Sec.~\ref{subsec:numerics_ops} is the operating envelope rather than a defect: at rate $3/4$ the codeword-confusion floor puts the silent-corruption target out of reach below $10$~dB, and within that envelope the shaping gain is what Fig.~\ref{fig:bar_D_vs_snr} reports; a lower-rate or stronger code would trade spectral efficiency for a lower frontier, which is precisely what link adaptation does. The frontier is thus a statement about this code at these \acp{snr}, not about the construction. The implementation that makes this geometry tractable replaces the $4$~s exhaustive inner maximization of \eqref{eq:LX} with a sum-product decoder plus systematic re-encoding ($17~\mu$s per decode, never overstating $L_{\mathcal{X}}$) and the coset classes and dual-code likelihood of Sec.~\ref{subsec:shaping_practical}, bringing a full operating point to under half an hour on a single processor core.

Second, the deployed point is resolved only to the accuracy of its threshold calibration. At $8$ and $10$~dB, where the silent-corruption cap is close to binding, a $20$-event calibration sets the threshold to a precision that the steep erasure curve turns into a factor of order $1.5$ in the reported erasure. The qualitative ordering there is robust, the third digit is not.

Third, $\bar D$ measures the observer's evidence per block in Stein's sense, and absolute values acquire operational meaning only against the observer's block budget. The averaging over Eve's prior deserves the same care: at high Bob \ac{snr} the mean is dominated by the upper tail of the log-normal, that is by realizations in which Eve's \ac{snr} is high enough to identify codewords outright whatever the cover, and against which no litter distribution offers protection. Reporting a worst case instead would need the observer's \ac{snr} to be bounded, which the shadowing model does not deliver and which an assumption about her position would have to supply.

Fourth, below $6$~dB the inner program of the relaxation is infeasible, because codeword confusion alone exceeds the silent-corruption cap, and the solver correctly returns uniform; the \ac{drl} policy, which cannot satisfy its penalty there, drifts up to $22\%$ above uniform. Neither is a property of shaped cover, and the regime is outside the operating envelope in any case, but in deployment infeasibility of \eqref{eq:shaping_convex} should be the signal to fall back to uniform rather than to the learned policy. Sec.~\ref{sec:conclusion} returns to these points.

\section{Conclusion}\label{sec:conclusion}

This paper formulated the litter-masked block z-channel: idle transmit slots carry sequences drawn from the codebook complement under a designable litter distribution, so that a public-design passive observer can no longer infer activity from the absence of a signal, while the upper layer keeps its erasure-only channel model. The Bayes-optimal \ac{map} decoder reduces to a single log-likelihood-ratio threshold, so receiver design, which minimizes erasure under a cap on silent corruption, is one-dimensional; transmitter-side litter shaping is a Stackelberg game with that threshold response as the inner problem, solved by an alternating convex relaxation and independently by a \ac{ppo} policy with Dirichlet action head and \ac{pid}-Lagrangian constraint handling. On a rate-$3/4$ $(16,12)$ instance in which the observer measures her own channel from the pilots but the transmitter knows it only through a log-normal prior with median $6$~dB below the legitimate link's, shaped litter reduces the observer's Stein detection exponent by $41$ to $48\%$ relative to uniform cover on the codebook complement. The reduction is flat in \ac{snr} above $10$~dB and flat in the activity rate, and the two solvers agree on it to within one standard error everywhere, which is the best evidence we have that the relaxation is not leaving much behind. Below $10$~dB a feasibility frontier set by codeword confusion, not by the cover distribution, leaves no operating point at all at this fixed coding rate, which is the region a rate-adaptive link would leave by lowering the rate. The cover is paid for in throughput at the frontier, about one percent of blocks at $10$~dB, and is free from $12$~dB upward, where holding the silent-corruption cap costs no measurable erasure.

Five directions remain open: extending the cover design from the $(16,12)$ geometry to standards-length codes, where the coset-class count $M^{n-k}-1$ itself grows large and calls for hierarchical litter-class structures; the design under a deployment constraint on the observer's position, which bounds her \ac{snr} and so replaces the expected exponent reported here by a worst case, the natural companion to the distributional statement this paper makes; the finite-blocklength information-theoretic companion study (achievability and converse bounds for the litter-masked block z-channel), deliberately scoped out here in favor of design; the outer-code view of Remark~\ref{rem:blockchannel}, in which the application-layer code rate over substitutions, erasures, and insertions replaces the silent-corruption cap as the design objective; and richer threat models, including multiple colluding observers, correlated large-scale variation of the legitimate and observing links, and online refinement of the look-up-table policy under channel-state drift.

\bibliographystyle{IEEEtran}
\bibliography{references}

\end{document}